\documentstyle[12pt, epsfig]{article}
\newcommand{\defprod}{\raisebox{-0.13cm}{~\shortstack{$\prod$ \\[-0.07cm] 
${}_{j\neq k}$}}~}

\def\tvi{\vrule height 12pt depth 6pt width 0pt}
\def\tv{\tvi\vrule}
\def\cc#1{\kern .7em\hfill #1 \hfill\kern .7em}
\begin{document}
\begin{flushright}
PM/01--14 \\
hep-ph/0103270
\end{flushright}
\begin{center}
{\Large \bf Comment on \\ ``Infrared Fixed Point Structure in Minimal 
Supersymmetric Standard Model with Baryon and Lepton Number Violation"}

\vspace{1cm}

{\sc  Y. Mambrini$^a$,  G. Moultaka$^b$ } 

\vspace{0.5cm}  {\it $^a$ CEA/DIF/DPTA/SPN \\
B.P. 12 F--91680 Bruy\`eres--le--Ch\^atel, France}

\vspace{0.3cm}
{\it  $^b$ Physique Math\'ematique et Th\'eorique, UMR No 5825--CNRS, \\
Universit\'e Montpellier II, F--34095 Montpellier Cedex 5, France.
}
\end{center}
\vspace{1cm}

\begin{abstract}
We reconsider the Infrared Quasi Fixed Points which were 
studied recently in the literature in the context of the 
Baryon and Lepton number violating Minimal Supersymmetric Standard Model
 (hep-ph/0011274). 
The complete analysis requires further care and reveals more structure than 
what was previously shown. The formalism we develop here
is quite general, and can be readily applied to a large class of models.  
\end{abstract}

\vspace{4cm}
\begin{center}
{\sl Submitted to Phys.Rev. D}
\end{center}

\newpage

Scenarios with R-parity violation, and in particular baryon or lepton number
violation within the Minimal Supersymmetric Standard Model 
($\not{\!\!R}_p$-MSSM),
have attracted attention in the past few years.
In a recent work, Ananthanarayan and Pandita have carried out 
a study of the Infrared Fixed Point structure of baryon/lepton number violating 
couplings in such scenarios \cite{pandita} .  Issues related to the Infrared Fixed 
Points, and in particular to the Infrared Quasi Fixed Points (IRQFP) of the 
running Yukawa type couplings, can provide generic information about 
the electroweak scale physics irrespectively of any high energy physics
assumptions. While IRQFPs are always present due to the Landau 
pole in the running of the Yukawa couplings, or equivalently due to a 
perturbativity bound on those couplings, exact fixed points  occur only in 
physically simplified configurations (projecting on some subset of Yukawa 
couplings and neglecting some gauge couplings). 
Of course when exact fixed points 
exist it is interesting to study their relation to the IRQFP as is done in 
\cite{pandita}. However one should keep in mind that such a relation
need not exist theoretically, not to mention that one expects anyway the IRQFP
regime to be physically more relevant than the one due to exact fixed points, 
as was initially shown in \cite{hill}. 
In the present comment we concentrate exclusively on the IRQFP
structure. In ref.\cite{pandita}, the general analytical forms
for the Yukawa couplings, derived in \cite{auberson}, were applied to the 
special case of the $\not{\!\!R}_p$-MSSM with the following (squared) running 
Yukawa couplings $\tilde{Y}_t, \tilde{Y}_b, \tilde{Y}_\tau, \tilde{Y}, 
\tilde{Y}', 
\tilde{Y}'' $ (see \cite{pandita} for notations). These analytical expressions
contain denominators of the form $1 + a_{jj} \tilde{Y}_j(0) \int_0^t (...)_j$.
The authors of \cite{pandita}
then claimed that ``{\sl in the regime where the Yukawa couplings 
$\tilde{Y}_t(0), \tilde{Y}_b(0), \tilde{Y}_\tau(0), \tilde{Y}(0), \tilde{Y}'(0)
, \tilde{Y}''(0) \to \infty $ with their ratios fixed, it is legitimate
to drop 1 in the denominators...}". This is in general a wrong statement.
``Neglecting 1" can happen to be justified in special cases, but this can
be seen only {\sl a posteriori} and at the expense of a more careful
study. For instance it was indeed proven to be correct in the R-parity conserving
MSSM ($R_p$-MSSM) in the $t/b/\tau$ sector \cite{kazakov}, but to be 
incorrect in the next to minimal $R_p$-MSSM \cite{mambrini}. The reason is
simple to understand from the analytical structure involved. 
The relevant Renormalization Group Equations (RGE) governing the running
of the Yukawa type couplings to one-loop order are of the form     
  

\begin{eqnarray}
\label{eq:rge}
\frac{d}{dt} \tilde{Y}_k(t) &=&\tilde{Y}_k(t)(\sum_ic_{ki} g^2_i(t)-\sum_la_{kl}\tilde{Y}_l(t)) 
\label{yukeq}
\end{eqnarray}

\noindent
where $t$ denotes the scale evolution parameter, $\tilde{Y}_k$ the squared 
Yukawa
couplings, $g^2_i$ the squared gauge couplings, and where $c_{ki}$ and $a_{kl}$
are constant coefficients depending on the model. 

\noindent
The general solution for such a system (valid for any number of Yukawa
couplings labeled by $k$) reads \cite{auberson}
\begin{equation}
\tilde{Y}_k(t)=\frac{\tilde{Y}_k(0) u_k(t)}{1+a_{kk}\tilde{Y}_k(0)\int_0^tu_k(t')dt'} 
\label{Ysol}
\end{equation}

\noindent
where the auxiliary functions $u_k$ are given by

\begin{equation}
u_k(t)= \frac{ E_k(t)}{\defprod ( 1 + a_{jj} \tilde{Y}_j(0) \int_0^t u_j(t')dt')^{a_{kj}/a_{jj}} }
\label{usol}
\end{equation}

\noindent
and the functions $E_k(t) \equiv exp[ \int_0^t \sum_ic_{ki} g^2_i(t')dt']$
are fully determined by the well-known running gauge couplings which need not
be written more explicitly here. In the case of \cite{pandita},
$k=1, ..., 6$ and the $u_k$'s correspond to $F_t, F_b, F_\tau, F, F', F''$
studied therein. It is easy to see from the structure of Eq.(\ref{usol})
that when some of (or all) the initial values $\tilde{Y}_k(0)\equiv Y^0 \to \infty$, 
 the $u_k$'s should have the form 
\begin{equation}
u_k \equiv \frac{u_k^{\mathrm{QFP}}}{{(Y^0)}^{p_k}} \label{uinfty}
\end{equation} 
where the $u_k^{\mathrm{QFP}}$'s
are initial condition independent functions,
$p_k \ge 0$ a set of values which depend on the model under consideration 
{\sl and on the choice of the subset of Yukawas having large initial conditions}. 
When for a given $j$, $\tilde{Y}_j(0) \to \infty$, it is obvious that ``1" 
can be safely dropped in 
$ 1 + a_{jj} Y^0_j \int_0^t u_j(t')dt'$ {\sl only if} $p_j < 1$,
 while if $p_j > 1$, ``1" becomes the leading contribution,
the $j^{th}$ sector then drops out completely from the expression of $u_k$
in (\ref{usol}) and the IRQFP value of $\tilde{Y}_j(t) \to 0$. It can also 
happen that $p_j=1$, in which case discarding ``1" requires the knowledge
of the actual
numerical contribution of the $Y^0$ independent term 
$a_{jj}\int_0^t u_j^{\mathrm{QFP}}$.

\noindent
We turn now to the determination of the powers $p_k$ in the case 
of $\not{\!\!R}_p$-MSSM. This is necessary 
for a correct study of the IRQFP regimes and has been overlooked
in ref. \cite{pandita}. We start off from Eqs.(18--23) of 
\cite{pandita}\footnote{ There are unfortunately some misprints in these
equations, the relevant ones for us are: (1) In Eq.(19), the power of the
third term in the denominator should read 1 instead of 1/6, 
(2) In Eq.(21) $\tilde{Y}'', F''$ should read $\tilde{Y}', F'$
(in their Eq.(8) $\lambda_{233}''$ should read $\lambda_{333}'$).}.

\noindent
Denoting by $p_t, p_b, p_\tau, p_0, p_1, p_2$ the powers corresponding
respectively to $F_t, F_b, F_\tau, F$, $F', F''$, one obtains the following
system,              

\begin{equation}
\left(
\begin{array}{l}
p_t \cr
p_b \cr
p_{\tau} \cr
p_0 \cr
p_1 \cr
p_2 \cr
\end{array}
\right)
=
\left(
\begin{array}{llllll}
0 & \frac{1}{6} & 0 &0 & \frac{1}{6} & \frac{1}{3}\cr
\frac{1}{6} & 0 & \frac{1}{4} & 0 & 1 &\frac{1}{3}\cr
0 & \frac{1}{2} & 0 & 1 & \frac{1}{2} &0\cr
0 & 0 &  1 & 0&  \frac{1}{2} &0 \cr
\frac{1}{6} &1& \frac{1}{4} & \frac{1}{4} & 0 & \frac{1}{3} \cr
\frac{1}{3}& \frac{1}{3}& 0 & 0 &\frac{1}{3}& 0 
\end{array}
\right)
\left(
\begin{array}{l}
(1 - p_t) \; \theta[1 - p_t] \; \delta_t \cr
(1 - p_b) \; \theta[1 - p_b] \; \delta_b \cr
(1 - p_{\tau}) \; \theta[1 - p_{\tau}] \; \delta_\tau \cr
(1 - p_0) \; \theta[1 - p_0] \; \delta_0 \cr
(1 - p_1) \; \theta[1 - p_1] \; \delta_1 \cr
(1 - p_2) \; \theta[1 - p_2] \; \delta_2 \cr
\end{array}
\right)
\label{matrice}
\end{equation}

\noindent
where the Heaviside $\theta$-function ($\theta( x_{ >0}, x_{<0}) =(1,0)$)
 accounts for the possibility of the 
$p_i$'s being larger or smaller than one, and $\delta_n$ 
($n=t, b, \tau, 0, 1, 2$)
takes the value $0 (1)$ when the corresponding initial $\tilde{Y}_n(0)$ is 
finite (infinite). We illustrate in table 1 the solutions of Eq.(\ref{matrice})
for three different sets of values of the $\delta_n$'s. In the first case
 $\delta_{t, b, 2}=1$, that is the initial conditions for only $\tilde{Y}_t,
 \tilde{Y}_b, \tilde{Y}''$ are taken infinitely large. This corresponds
to the case studied in \cite{pandita} where $\tilde{Y}_{\tau}, \tilde{Y}, 
\tilde{Y}'$ were put to zero in order to study the exact (attractive) IR 
fixed point. As one can see from the table, all the powers but $p_1(=1)$
turn out to be strictly smaller than one, so that "1'' can indeed be
safely neglected in the corresponding contributions. Moreover, 
since in the case we 
consider $\tilde{Y}'(0)$ is not taken infinitely large, the contribution
of $F'$ will drop out anyway, being suppressed by a factor $1/(Y^0)^{p_1}$ where
$Y^0$ is the large initial condition. So in retrospect, the legitimacy of
dropping "1'' in this case was a lucky situation. [The reason is that
the reduced matrix of the system $p_t, p_b, p_\tau$ satisfies a sufficient
condition to forbid $p_i >1$, namely that the sum of the matrix elements in 
each row is less than 1.] We thus agree with the authors of 
ref. \cite{pandita} on this part of their study, including the numerical
estimate of the IRQFP values. 
The point is that such configurations do not prevail for 
more general initial conditions. In the second case
of table 1, where we keep finite either $\tilde{Y}_\tau(0)$ or $\tilde{Y}'(0)$
(but not necessarily both), $p_\tau, p_1$
are found to be both greater than one. Clearly here, neglecting
one in Eq.(\ref{usol}) in the $\tau$ sector, when 
$\tilde{Y}_\tau(0) \to \infty$, (or in the ``prime" sector when
$\tilde{Y}'(0) \to  \infty$) leads to a wrong result! Moreover,
$p_\tau, p_1 >1$ has a physically interesting consequence: 
the IRQFP values $\tilde{Y}_\tau^{\mathrm{QFP}}(t),
\tilde{Y}'^{\mathrm{QFP}}(t)$ are zero, as can been seen from 
Eqs.(\ref{Ysol}, \ref{uinfty}). This is at variance with what Eq.(39) of 
ref.\cite{pandita} would have implied.
There is thus a much richer structure than stated in \cite{pandita},
in the sense that the Yukawa couplings in the IRQFP regime can be 
 naturally driven to the hyperplane which has the exact infrared fixed points
without a prior projection on this hyperplane as an initial condition. 
This behavior is actually more general than in the case $\delta_{t, b, 2}=1$.
It holds even when all initial conditions are taken infinite 
($\delta_{all}=1$). The latter case, considered in \cite{pandita},
is even more tricky to handle, since it leads to two different sets of solutions for Eq.(\ref{matrice}) as can be seen from table 1. These solutions 
correspond to two different behaviors in the IRQFP limit. However, 
due to the uniqueness of the running Yukawa's for given initial 
conditions (no singularities are encountered when running from the GUT scale 
down to the electroweak scales), one of these solutions should be spurious. 
The correct solution turns out to be
{\bf (1)}. This is found consistently, both from analytical considerations 
which we do not describe here, and from a numerical study illustrated in 
figure 1.  Solution {\bf (2)} would have required all the Yukawa couplings
to become $\tilde{Y}^0$ independent when the latter goes increasingly large,
while in solution {\bf (1)} the couplings  $\tilde{Y}'$
and $\tilde{Y}_\tau$ are expected to {\sl decrease} to zero. 
Figure 1 exhibits clearly the features of solution {\bf (1)}, 
where the logs of $\tilde{Y}'$ and $\tilde{Y}_\tau$ decrease linearly
as expected and 
with rates close to the expected ones (respectively  $.25$ and $.35$).
The behavior of $\tilde{Y}'^{\mathrm{QFP}}$ and 
$\tilde{Y}_\tau^{\mathrm{QFP}}$
contradicts Eq.(39) of ref. \cite{pandita}. Moreover, 
the fact that $p_\tau$ and $p_1$ are greater than one means that it is the
$F'^{\mathrm{QFP}}$ and $F_\tau^{\mathrm{QFP}}$ contributions 
(and not the associated ``1"'s !) that should be
dropped out in the right-hand side of Eqs.(40)--(45) of ref. \cite{pandita}.

\noindent
Finally, we note that comments similar to the above, hold as well in 
relation to the recent work 
\cite{pandita1}.

\begin{table}
$$\vbox{\offinterlineskip \halign{
\tv# & \cc{#} & \tv# & \cc{#}  & \tv# & \cc{#} & \tv# & \cc{#} & \tv# &\cc{#} & \tv# & \cc{#} & \tv# & \cc{#} & \tv#\cr
\noalign{\hrule}
&     &&\cc{$p_t$}    &&\cc{$p_b$}   &&\cc{$p_{\tau}$} &&\cc{$p_0$}      && \cc{$p_1$}    && \cc{$p_2$}   &\cr
\noalign{\hrule}
\hline
\noalign{\hrule}
&  $\delta_{t, b, 2}=1$   &&\cc{$5/17$} && \cc{$5/17$} && \cc{$6/17$}  && \cc{$0$} && \cc{$1$}      && \cc{$8/17$}&\cr
\noalign{\hrule}
\hline
\noalign{\hrule}
& $\delta_{\tau \; or \; 1}=0$   &&\cc{$5/17$}    &&\cc{$5/17$}    &&\cc{$23/17$}   && \cc{$0$}   && \cc{$5/4$}  && \cc{$8/17$}  &\cr
\noalign{\hrule}
\hline
\noalign{\hrule}
& {\bf (1)} $\delta_{all}=1$   &&\cc{$5/17$}   && \cc{$5/17$} &&\cc{$23/17$}    && \cc{$0$}    &&\cc{$5/4$} && \cc{$8/17$}  &\cr
\noalign{\hrule}
& {\bf (2)} $\delta_{all}=1$   &&\cc{$41/146$} && \cc{$1$} && \cc{$18/73$}  && \cc{$1$} && \cc{$37/173$}      && \cc{$59/146$}&\cr
\noalign{\hrule}
\hline
\noalign{\hrule}
}}$$
\caption{solutions for the powers in three different IRQFP regimes.}
\end{table}

\newpage
\vspace{-12cm}

\begin{figure}[htb]
\begin{center}
\psfig{figure=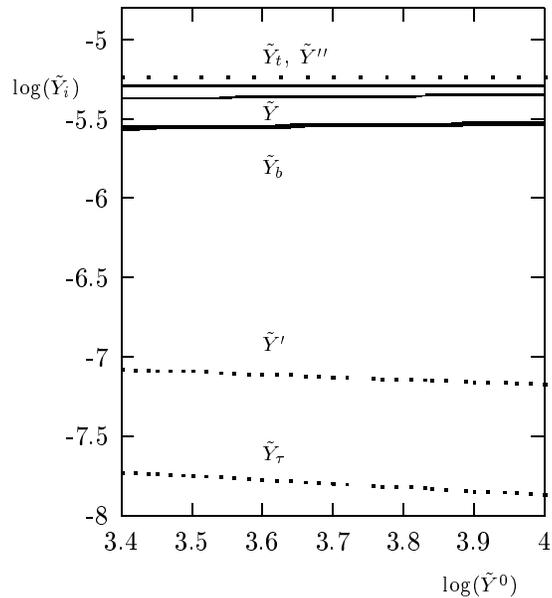}
\end{center}
\vspace{-14cm}
\caption{Logs of the various Yukawa couplings $\tilde{Y}_k(t)$ at $t=1$ TeV, 
as a function of  their common initial condition Log $(\tilde{Y}^0)$, where $\tilde{Y}^0$ takes large values 
in the range 30--60. Solution {\bf (1)} is clearly singled out.}       
\end{figure} 

\begin{thebibliography}{99}
\bibitem{pandita} B. Ananthanarayan, P. N. Pandita, hep-ph/0011274,
{\it Phys. Rev.} {\bf D63} (2001) 076008.
\vspace{-.3cm}
\bibitem{hill} C. T. Hill, Phys. Rev. {\bf D24} (1981)  691.
\vspace{-.3cm}
\bibitem{auberson} G. Auberson, G. Moultaka, {\it Eur. Phys. J.} {\bf C12}
(2000) 331.
\vspace{-.3cm}
\bibitem{kazakov} D.I. Kazakov, G. Moultaka, {\it Nucl. Phys.} {\bf B577}
121 (2000).
\vspace{-.3cm}
\bibitem{mambrini} Y. Mambrini, G. Moultaka, M. Rausch de Traubenberg,
hep-ph/0101237.
\vspace{-.3cm}
\bibitem{pandita1} P. N. Pandita, hep-ph/0103005.
\end{thebibliography}
\end{document}